\documentclass{article}
\usepackage{arxiv}

\usepackage[utf8]{inputenc} 
\usepackage[T1]{fontenc}    
\usepackage{hyperref}       
\usepackage{url}            
\usepackage{booktabs}       
\usepackage{microtype}      
\usepackage{lipsum}	    	
\usepackage{graphicx}
\usepackage{doi}
\usepackage[section]{placeins}
\usepackage{mathtools}
\usepackage{amsmath}
\usepackage{amssymb}
\usepackage[ruled,vlined]{algorithm2e}

\title{\textbf{Enhanced physics-informed neural networks for hyperelasticity}}

\author{{\hspace{1mm}Diab W. Abueidda}\thanks{abueidd2@illinois.edu} \\
	National Center for Supercomputing Applications\\
	Department of Mechanical Science and Engineering\\
	University of Illinois at Urbana-Champaign\\
	\And
	{\hspace{1mm}Seid Koric}\\
	National Center for Supercomputing Applications\\
	Department of Mechanical Science and Engineering\\
	University of Illinois at Urbana-Champaign\\
	\And
	{\hspace{1mm}Erman Guleryuz}\\
	National Center for Supercomputing Applications\\
	University of Illinois at Urbana-Champaign\\
	\And
	{\hspace{1mm}Nahil A. Sobh}\\
	Center for Artificial Intelligence Innovation\\
	National Center for Supercomputing Applications\\
	University of Illinois at Urbana-Champaign\\
}

\renewcommand{\shorttitle}{Enhanced PINNs for hyperelasticity}

\begin{document}

\maketitle

\begin{abstract}
Physics-informed neural networks have gained growing interest. Specifically, they are used to solve partial differential equations governing several physical phenomena. However, physics-informed neural network models suffer from several issues and can fail to provide accurate solutions in many scenarios. We discuss a few of these challenges and the techniques, such as the use of Fourier transform, that can be used to resolve these issues. This paper proposes and develops a physics-informed neural network model that combines the residuals of the strong form and the potential energy, yielding many loss terms contributing to the definition of the loss function to be minimized. Hence, we propose using the coefficient of variation weighting scheme to dynamically and adaptively assign the weight for each loss term in the loss function. The developed PINN model is standalone and meshfree. In other words, it can accurately capture the mechanical response without requiring any labeled data. Although the framework can be used for many solid mechanics problems, we focus on three-dimensional (3D) hyperelasticity, where we consider two hyperelastic models. Once the model is trained, the response can be obtained almost instantly at any point in the physical domain, given its spatial coordinates. We demonstrate the framework's performance by solving different problems with various boundary conditions.
\end{abstract}


\section{Introduction} \label{intro}

Many physical phenomena can be described by partial differential equations (PDEs). Nevertheless, analytical solutions are generally limited to those with simple geometries and loading conditions. Thus, several numerical methods have been developed to approximate solutions. Conventional computational methods utilized to solve PDEs involved in solid mechanics problems are the mesh-free methods (\cite{huerta2018meshfree}), finite element analysis (FEA) (\cite{hughes2012finite}), and isogeometric analysis (\cite{hughes2018isogeometric}). Despite the fact that these numerical methods have proven successful for many engineering problems, they have encountered several challenges for specific problems such as ill-posed, high-dimensional, and coupled problems.

Recently, rapid growth in the utilization of deep learning (DL) and data-driven modeling has been seen in computational solid mechanics \cite{kim2021deep, shahane2022surrogate, rong2019predicting, mozaffar2019deep, koric2021deep, fatehi2021accelerated, spear2018data, he2022exploring, gu2018bioinspired, linka2021constitutive}. While deep learning offers a platform for rapid inference, it needs a large number of data points to learn the multifaceted correlations between the outputs and inputs. The number of data points required to train a model is problem-based, i.e., intricate problems require large datasets to produce models able to capture the response accurately. Accordingly, such surrogate deep learning models usually mandate a discretization method, such as the finite element method, to generate the data required to train the model, while data generation is usually the bottleneck step in developing any data-driven model.

Another direction in which deep learning is used to capture physical phenomena is physics-informed neural networks (PINNs), which are of eminent interest to industry and academia \cite{raissi2019physics, abueidda2021meshless, haghighat2021physics, cai2021physics}. PINNs have been used to approximate solutions for boundary value problems without necessitating the use of labeled data due to their abilities as universal approximators \cite{hornik1989multilayer}. PINNs have been utilized in different ways. For example, one approach in which PINNs are used is to define the loss function as the PDEs' residual at specific collocation points in the physical domain and on its corresponding boundary and initial conditions \cite{henkes2022physics, niaki2021physics, rao2021physics}. This approach is commonly called the deep collocation method (DCM). Additionally, Samaniego et al. \cite{samaniego2020energy}, Nguyen-Thanh et al. \cite{nguyen2020deep, nguyen2021parametric}, and Abueidda et al. \cite{abueidda2022deep} presented a deep energy method (DEM), where the loss function is defined in terms of the potential energy to solve various problems such as elasticity, hyperelasticity, viscoelasticity, and piezoelectricity. Such a loss function reduces the prerequisites on the differentiability of the basis function, and it automatically satisfies traction-free boundary conditions. 

Fuhg et al. \cite{fuhg2022mixed} have demonstrated that both the deep collocation method and deep energy method suffer to resolve fine displacement and stress features.Consequently, they have proposed a mixed deep energy method (mDEM), in which stress is an additional output of the neural network. Then, an additional loss term is added to the introduced loss function to account for the difference between the stress obtained directly from the neural network and the stress calculated using the displacement field. They have shown that such a formulation enhances the capability of the neural network to capture localized features in finite strain hyperelasticity.

Even though PINNs have apparent success in solving various boundary value problems, little is comprehended about how such neural networks behave during the optimization process and why those models sometimes fail to train. Additionally, the training of PINNs can be slow and challenging, and it might mandate a substantial effort to devise the network's architecture and fine-tune its hyperparameters, which are often obtained in a trial and error fashion or grid search technique. Grid search technique can be prohibited if there are many hyperparameters. There are several attempts to understand possible failure modes in PINNs. Krishnapriyan et al. \cite{krishnapriyan2021characterizing} argue that minimizing the residual of the PDE system as a soft regularization in the loss function may be easier to deploy, but this comes with trad-offs; in many cases, the optimization problem becomes very challenging to solve. Nevertheless, they have illustrated that neural networks have the expressivity/capacity to learn and approximate solutions if specific considerations are taken into account. For instance, they have proposed using curriculum PINN regularization as a potential technique to enhance PINNs' performance \cite{bengio2009curriculum}. Additionally, Fuks et al. \cite{fuks2020limitations} have shown that PINNs provide accurate approximate solutions of hyperbolic PDEs in the absence of discontinuities. They fail in the presence of shocks, while adding a diffusion term can recover the proper location and size of the shock.   

Another source of complexity in PINNs is the multi-term loss function. Wang et al. \cite{wang2021understanding} studied the distribution of the gradient of each term in the loss function, and they proposed to use a gradient scaling approach for determining the weights of the loss terms. Wang et al. \cite{wang2022and} suggested another approach for specifying the weights of the loss terms that is based on the eigenvalues of the neural tangent kernel (NTK) matrix. Moreover, Tancik et al. \cite{tancik2020fourier} have indicated that passing input points through a simple Fourier feature mapping helps a multilayer perceptron (MLP) learn high-frequency functions in low-dimensional problem domains. Based on the neural tangent kernel (NTK) theory, they have demonstrated that a standard MLP fails to learn high frequencies in theory and practice. They have utilized a Fourier feature mapping to overcome this spectral bias to transform the effective NTK into a stationary kernel with a tunable bandwidth. Wang et al. \cite{wang2021eigenvector} have extended the work of Tancik et al. \cite{tancik2020fourier} to PINNs, examined the limitation of PINNs via the neural tangent kernel theory, and illuminated how PINNs are biased towards learning functions along the dominant eigen-directions of their limiting NTK. Also, they have shown that using Fourier feature mappings with  PINNs can modulate the frequency of the NTK eigenvectors.

Groenendijk et al. \cite{groenendijk2021multi} proposed a weighting scheme based on the coefficient of variations, where the set of weights depends on properties observed while training the model. The proposed method assimilates a measure of uncertainty to balance the losses, and consequently, the loss weights evolve during training without demanding another optimization. Additionally, they have illustrated that coefficient of variation (CoV) weighting outperforms other weighting schemes such as GradNorm \cite{chen2018gradnorm}. When a physics-informed neural network is developed, one ends with a loss function defined as a weighted linear combination of multiple losses, where the final performance is sensitive to the weights for these losses. 

In this paper, we are address the approximate solutions of hyperelasticity problems, where an enhanced PINN model is proposed. The output layer of the neural network consists of both displacement and stress components. We define our loss function as the sum of loss terms appearing in the DCM and DEM. Then, an additional loss term is added, accounting for the difference between the stresses calculated using the displacement profile and those predicted directly from the neural network. Accordingly, our loss function has many terms. Thus, we adopt the weighting scheme proposed by Groenendijk et al. \cite{groenendijk2021multi} and perform the first implementation of CoV weighting in the context of physics-informed neural networks. Furthermore, Gaussian Fourier feature mapping and curriculum learning are implemented to improve the network's performance. 

The paper's outline is as follows: Section \ref{hyper} details the governing equations of the hyperelasticity problem. Section \ref{method} presents the details of the approach, where a general problem setup is discussed. Then, in Section \ref{results}, the enhanced PINNs are used to solve three-dimensional (3D) examples. Finally, we conclude the paper in Section \ref{conclu} by stating the important results and highlighting possible future work directions.
\section{Hyperelasticity} \label{hyper}

A body made of a isotropic and homogeneous hyperelastic material under finite deformation is considered. The mapping $\boldsymbol{\zeta}$ of material points from the initial configuration to the current configuration is expressed as:
\begin{equation}\label{mapping}
\begin{aligned}
    \boldsymbol{x}&=\boldsymbol{\zeta}\left(\boldsymbol{X},t\right)=\boldsymbol{X}+\boldsymbol{\hat{u}}.\\
\end{aligned}
\end{equation}
Assuming that inertial forces are absent, the strong form is determined by:
\begin{equation}\label{eqlbrm_HE}
\begin{split}
    \boldsymbol{\nabla}_{\boldsymbol{X}}\cdot \boldsymbol{P} + \boldsymbol{f}_B &=\boldsymbol{0}{,} \quad  \boldsymbol{X}\in\Omega,\\
    \boldsymbol{u}&=\overline{\boldsymbol{u}}, \quad \! \boldsymbol{X}\in\Gamma_{u},\\
    \boldsymbol{P}\cdot\boldsymbol{N}&=\overline{\boldsymbol{t}}, \quad  \boldsymbol{X}\in\Gamma_{t},\\
\end{split}
\end{equation}
where $\boldsymbol{P}$ is the first Piola-Kirchhoff stress, $\boldsymbol{f}_B$ is the body force, $\boldsymbol{\nabla}_{\boldsymbol{X}} \cdot$ is the divergence operator, and $\boldsymbol{N}$ represents the outward normal unit vector in the initial configuration. $\overline{\boldsymbol{t}}$ denotes a defined natural boundary condition, and $\overline{\boldsymbol{u}}$ accounts for a defined essential boundary condition. $\Omega$ denotes the material domain, while $\Gamma_u$ and $\Gamma_t$ are the surfaces with essential and natural boundary conditions, respectively. The constitutive law for such a material is written as: 
\begin{equation}\label{constit_hyper}
\begin{aligned}
    \boldsymbol{P}&=\frac{\partial \psi\left(\boldsymbol{F}\right)}{\partial \boldsymbol{F}}\\
    \boldsymbol{F}&=\boldsymbol{\nabla}_{\boldsymbol{X}} \boldsymbol{\zeta}\left(\boldsymbol{X}\right)\\
\end{aligned}
\end{equation}
where $\boldsymbol{F}$ presents the deformation gradient, and $\psi$ denotes the strain energy density of a specific material. Although the framework discussed is not limited to specific material, in this paper, we focus on materials obeying two hyperelastic models: Neo-Hookean and Lopez-Pamies models \cite{lopez2010new}. For the neo-Hookean hyperelastic material, the strain free energy $\psi\left(\boldsymbol{F}\right)$ is given by:
\begin{equation}\label{neohookenergy}
\begin{aligned}
    \psi\left(\boldsymbol{F}\right)&=\frac{1}{2}\;\lambda\; \left(\text{ln}\left(J\right)\right)^{2}-\mu \; \text{ln}\left(J\right)+\frac{1}{2} \; \mu \; \left(I_{1} -3\right),\\
\end{aligned}
\end{equation}
where the first principal invariant is determined by $I_{1} = \text{trace} \left(\boldsymbol{C} \right)$, the right Cauchy-Green tensor $\boldsymbol{C}$ is defined as $\boldsymbol{C} =\boldsymbol{F}^{T} \boldsymbol{F}$, where $\left(\right)^{T}$ is the transpose operator. $J$ is the determinant of the deformation gradient, $J=\text{det}\left(\boldsymbol{F}\right)$. Using Equation (\ref{constit_hyper}), $\boldsymbol{P}$ is given by:
\begin{equation}\label{constit_neo}
\begin{aligned}
    \boldsymbol{P}&=\frac{\partial \psi\left(\boldsymbol{F}\right)}{\partial \boldsymbol{F}}=\mu \; \boldsymbol{F}+\left(\lambda\;\text{ln}\left(J\right)-\mu \right)\;\boldsymbol{F}^{-T}.\\
\end{aligned}
\end{equation}
For Lopez-Pamies hyperelastic model, the strain energy model is expressed as:
\begin{equation}\label{strain_energy_hyper}
\begin{aligned}
    \psi &= \sum_{r=1}^{M} \frac{3^{1-\alpha_r}}{2 \alpha_r} \mu_r \left(I_1^{\alpha_r} - 3^{\alpha_r} \right) - \sum_{r=1}^{M} \mu_{r} \text{ln} J + \frac{\lambda}{2} \left(J-1\right)^{2}\\
\end{aligned}
\end{equation}
where $M$ represents the number of terms included in the summation, while $\alpha_r$, $\mu_r$ and $\lambda$ are material constants $\left( r=1,2,..., M\right)$. Using Equation (\ref{constit_hyper}), $\boldsymbol{P}$ is given by:
\begin{equation}\label{stress_hyper}
\begin{aligned}
    \boldsymbol{P} &= \frac{\partial \psi\left(\boldsymbol{F}\right)}{\partial \boldsymbol{F}} = \sum_{r=1}^{M} 3^{1-\alpha_r}\mu_r I_1^{\alpha_r-1} \boldsymbol{F} - \sum_{r=1}^{M} \mu_{r} \boldsymbol{F}^{-T} + \lambda \left(J^{2}-J\right) \boldsymbol{F}^{-T}.\\
\end{aligned}
\end{equation}
In this paper, the number of terms $M$ utilized to describe the material's response, see Equation (\ref{strain_energy_hyper}) and (\ref{stress_hyper}), is $M=2$. Regardless the material model used, the energy functional can be minimized with regards to $\boldsymbol{u}$ to obtain the deformation that fulfills static equilibrium. Neglecting inertial forces and considering only conservative loads, the potential energy functional of the body is expressed as: 
\begin{equation}\label{hyper_weak}
\begin{aligned}
    \Pi = \underbrace{\int_{\Omega} \psi\left(\boldsymbol{u}\right) d\Omega}_{\text{internal energy}} \underbrace{- \int_{\Omega} \boldsymbol{u}^{T}\boldsymbol{f}_{b} d\Omega - \int_{\Gamma} \boldsymbol{u}^{T} \overline{\boldsymbol{t}} d\Gamma}_{\text{external energy}}.
\end{aligned}
\end{equation}
The Cauchy stress $\boldsymbol{S}$ can be calculated from the first Piola-Kirchhoff stress using:
\begin{equation}\label{Cauchy}
\begin{aligned}
    \boldsymbol{S} &= \frac{1}{J} \boldsymbol{P} \boldsymbol{F}^{T}.\\
\end{aligned}
\end{equation}
\section{Methods} \label{method}

The finite element method (FEM) is typically employed to solve problems with geometric and/or material nonlinearities. Specifically, when an implicit finite element scheme is utilized with an iterative method (e.g., Newton-Raphson), the tangent matrix and residual vector are assembled and then exploited to solve the linear system of equations using an iterative or direct solver to find the unknowns in each iteration. On the contrary, an explicit nonlinear finite element scheme does not solve a linear system of equations. Nonetheless, it usually requires small time increments, and conditional stability often limits it. Moreover, when explicit FEM is utilized for quasi-static simulations, one must ensure that the inertial forces are negligible \cite{koric2009explicit}.

This paper attempts to use an alternative numerical scheme to solve such problems. We employ deep learning to approximate the displacement and stress profiles. Here, the approach is meshfree, where the loss function is minimized within a deep learning framework such as PyTorch \cite{NEURIPS2019_9015}. The proposed loss function consists of multiple terms, including potential energy and residuals of the strong form, where we use CoV weighting to automatically determine the coefficients of the different loss terms.

\subsection{Deep feedforward neural networks}\label{NN}

\begin{figure}[!htb]
    \centering
    \includegraphics[width=0.65\textwidth]{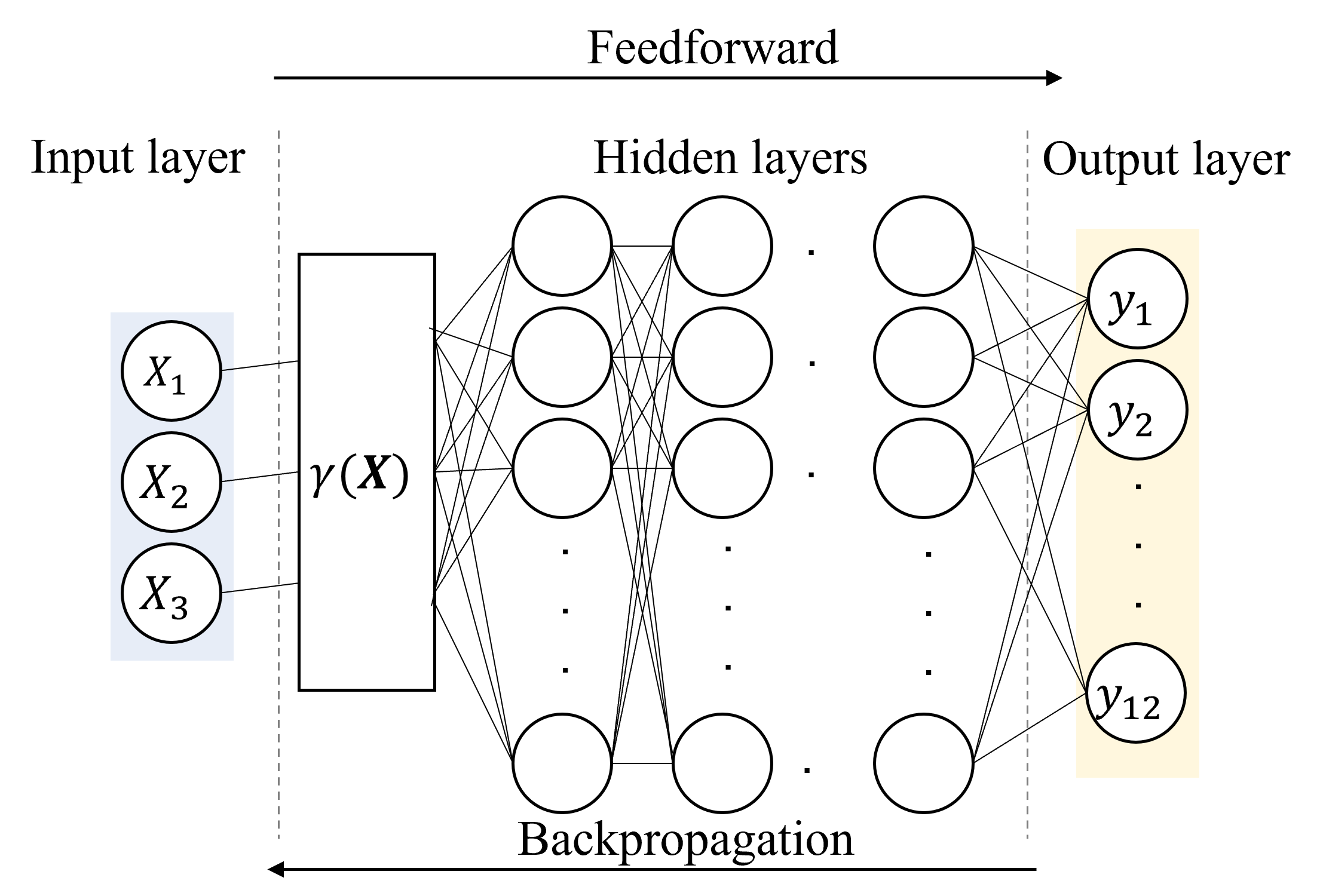}
    \caption{Fully connected (dense) artificial neural network with Fourier feature mapping.}
    \label{DANN}
\end{figure}

Deep feedforward neural networks are layers of interlinked individual unit cells, named neurons, connected to other neurons’ layers. Figure \ref{DANN} displays the deep feedforward neural network, composed of linked layers of neurons that compute an output layer (predictions) given input data. The layers of neurons propagate information forward to the subsequent layers, assembling a learning network with some feedback mechanism. The depth of a neural network is measured by how many hidden layers, layers between output and input layers.

\FloatBarrier

The layers are connected through weights $\boldsymbol{W}$ and biases $\boldsymbol{b}$. For a layer $l$, the output $\boldsymbol{\hat{Y}}^{l}$ is written as:
\begin{equation}\label{AEq1}
\begin{aligned}
    \boldsymbol{Z}^{l}&=\boldsymbol{W}^{l} \boldsymbol{\hat{Y}}^{l-1}+\boldsymbol{b}^{l}\\
    \boldsymbol{\hat{Y}}^{l}&=f^{l}\left(\boldsymbol{Z}^{l}\right)\\
\end{aligned}
\end{equation}
where the weights $\boldsymbol{W}$ and biases $\boldsymbol{b}$ are updated after every training (optimization) iteration. The activation function $f^{l}$ is an $\mathbb{R} \rightarrow \mathbb{R}$ mapping that transforms vector $\boldsymbol{Z}^{l}$, calculated using weights and biases, into output for every neuron in the layer $l$. Activation functions are nonlinear functions such as rectified linear unit (ReLu), sigmoid, and hyperbolic tangent. They help the neural network learn nearly any intricate functional correlation between outputs and inputs. 

Tancik et al. \cite{tancik2020fourier} have shown that passing the input layer through a simple Fourier feature mapping promotes a multilayer perceptron to learn high-frequency functions in low-dimensional problem domains. Here, the input layer (coordinates of the sampled points from the physical interest) is passed through a Fourier feature mapping $\gamma$ before passing them through the multilayer perceptron. The function $\gamma$ maps input points $\boldsymbol{X}$ to the surface of higher-dimensional hypersphere using a set of sinusoids:
\begin{equation}\label{Fourier}
\begin{aligned}
    \gamma\left(\boldsymbol{X}\right) &= \left[ a_1 cos\left(2\pi\boldsymbol{h}_{1}^{T} \boldsymbol{X} \right), a_1 sin\left(2\pi\boldsymbol{h}_{1}^{T} \boldsymbol{X} \right),...,a_m cos\left(2\pi\boldsymbol{h}_{m}^{T} \boldsymbol{X} \right), a_m sin\left(2\pi\boldsymbol{h}_{m}^{T} \boldsymbol{X} \right)\right]\\
\end{aligned}
\end{equation}
where $\boldsymbol{h}_{i}$ are the Fourier basis frequencies used for approximation, and $a_{i}$ denote the corresponding Fourier series coefficients. Here, we adopt Gaussian random Fourier features (RFF). Specifically, each entry in $\boldsymbol{h}$ is sampled from a normal distribution $\mathcal{N}\left(0,\sigma^2\right)$, where $\sigma^2$ is the variance. $\sigma$ is a hyperparameter that has to be specified for each problem. 
 
The loss function $\mathcal{L}$ computes a loss value that reveals how well the network’s predictions compare with targets after each feedforward pass. The loss function is utilized to get the weights $\boldsymbol{W}$ and biases $\boldsymbol{b}$ inducing a minimized loss value. The process of attaining the optimized weights and biases in machine learning is called training. Backpropagation is employed throughout the training process, where the loss function is minimized iteratively. One of the most common and most straightforward optimizers used is gradient descent \cite{pattanayak2017pro}:
\begin{equation}\label{AEq2}
\begin{aligned}
    W_{ij}^{c+1}&=W_{ij}^{c}-\beta \frac{\partial \, \mathcal{L}}{\partial W_{ij}^{c}}\\
        b_{i}^{c+1}&=b_{i}^{c}-\beta \frac{\partial \, \mathcal{L}}{\partial b_{i}^{c}}\\
    \end{aligned}
\end{equation}

where $\beta$ denotes the learning rate. Equation (\ref{AEq2}) displays the formula utilized to update the weights $\boldsymbol{W}$ and biases $\boldsymbol{b}$ at a given iteration $c$ within the gradient descent training process. The details of the loss function used in this paper are discussed below.

\subsection{Physics-informed neural networks}\label{PINNs}

\begin{figure}[!htb]
    \centering
    \includegraphics[width=1.0\textwidth]{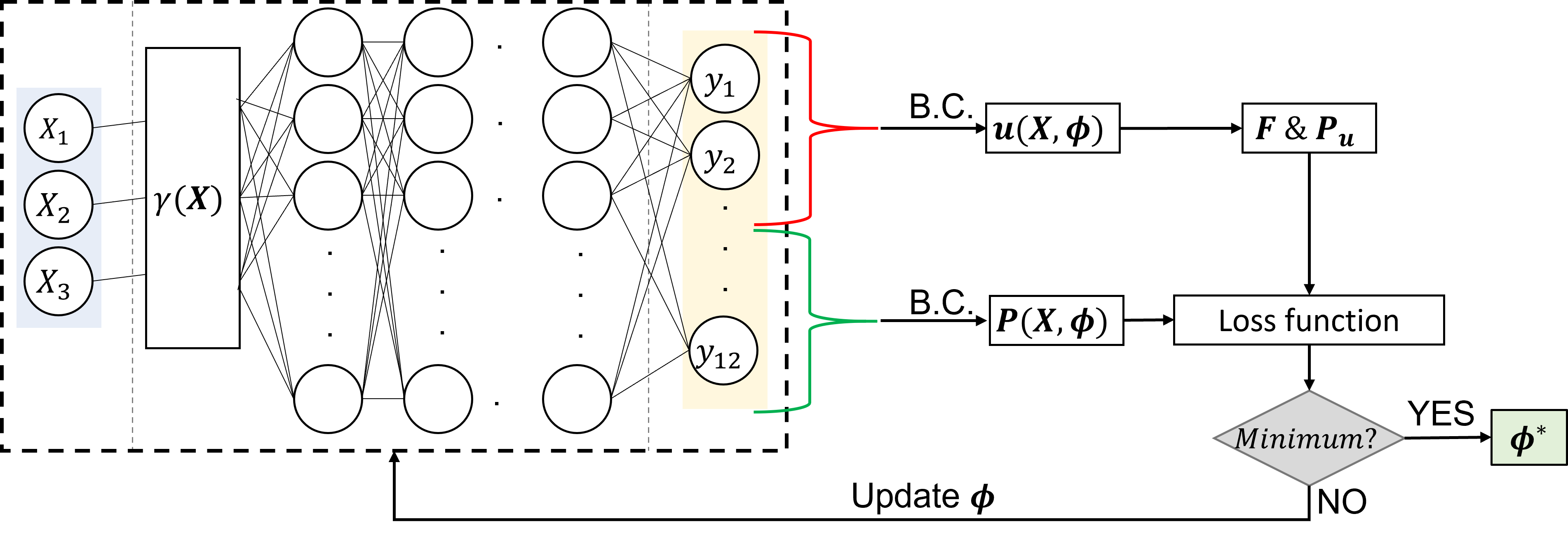}
    \caption{Flow chart of the proposed PINNs.}
    \label{flowchart}
\end{figure}

This section discusses how PINNs are used to solve hyperelasticity problems. The loss function, consisting of multiple loss terms, is minimized using deep learning tools, as depicted in Figure \ref{flowchart}. The approximate solution $\boldsymbol{y} \left(\boldsymbol{X}; \boldsymbol{\phi} \right)$ of the governing equations is obtained by training a neural network with parameters $\boldsymbol {\phi}= \{\boldsymbol{W}, \boldsymbol{b}\}$. Specifically, the deep neural network minimizes the loss function $\mathcal{L}$ to obtain the optimized network parameters $\boldsymbol{\phi}^{*} =\{\boldsymbol{W}^*, \boldsymbol{b}^*\}$, where the neural network is used as global shape function for the output through the physical domain of interest. The neural network maps the coordinates $\boldsymbol{X}$ of the sampled points to an output $\boldsymbol{y}\left(\boldsymbol{X}, \boldsymbol{\phi}\right)$ employing the feedforward propagation. $\boldsymbol{X}$ is composed of three components, representing the three spatial coordinates, while $\boldsymbol{y}$ has 12 components, three for the displacement vector and nine for the stress tensor.

There are two approaches to fulfill the boundary conditions. The first approach is to subject $\boldsymbol{y}\left(\boldsymbol{X}, \boldsymbol{\phi}\right)$ to:
\begin{equation}\label{NN_BC1}
\begin{aligned}
    \boldsymbol{u}\left(\boldsymbol{X}, \boldsymbol{\phi}\right) &= \boldsymbol{A}\left(\boldsymbol{X}\right) + \boldsymbol{B}\left(\boldsymbol{X}\right) \circ \boldsymbol{y}_{u}\left(\boldsymbol{X}, \boldsymbol{\phi}\right)\\   
    \boldsymbol{P}\left(\boldsymbol{X}, \boldsymbol{\phi}\right) &= \boldsymbol{C}\left(\boldsymbol{X}\right) + \boldsymbol{D}\left(\boldsymbol{X}\right) \circ \boldsymbol{y}_{P}\left(\boldsymbol{X}, \boldsymbol{\phi}\right)\\   
\end{aligned}
\end{equation}
where $\boldsymbol{A}\left(\boldsymbol{X}\right)$, $\boldsymbol{B}\left(\boldsymbol{X}\right)$, $\boldsymbol{C}\left(\boldsymbol{X}\right)$, and $\boldsymbol{D}\left(\boldsymbol{X}\right)$ are distance functions chosen such that $\boldsymbol{u}\left(\boldsymbol{X}, \boldsymbol{\phi}\right)$ and  $\boldsymbol{P}\left(\boldsymbol{X}, \boldsymbol{\phi}\right)$ satisfy the boundary conditions active on the physical domain. $\boldsymbol{y}_{u}\left(\boldsymbol{X}, \boldsymbol{\phi}\right)$ and $\boldsymbol{y}_{P}\left(\boldsymbol{X}, \boldsymbol{\phi}\right)$ are the outputs of the neural network before the enforcement of boundary conditions, corresponding to the displacement and stress, respectively. More details can be found in the work of Nguyen-Thanh et al. \cite{nguyen2020deep} and Rao et al. \cite{rao2021physics}. The second approach to fulfill the boundary conditions is to set:
\begin{equation}\label{NN_BC2}
\begin{aligned}
    \boldsymbol{u}\left(\boldsymbol{X}, \boldsymbol{\phi}\right) &=  \boldsymbol{y}_u\left(\boldsymbol{X}, \boldsymbol{\phi}\right)\\
    \boldsymbol{P}\left(\boldsymbol{X}, \boldsymbol{\phi}\right) &=  \boldsymbol{y}_P\left(\boldsymbol{X}, \boldsymbol{\phi}\right)\\
\end{aligned}
\end{equation}
where the boundary conditions are accounted for through the optimization process of weights and biases of the deep neural network by adding an extra term to the loss function. This is commonly known as soft enforcement of boundary conditions. Here, the former approach is employed for the displacement boundary conditions. At the same time, the latter is used for the traction boundary conditions.

Since the neural network predicts both displacement and stress, training of the constitutive law is required. In other words, we set the first Piola-Kirchhoff stress $\boldsymbol{P}\left(\boldsymbol{X}, \boldsymbol{\phi}\right)$ predicted using the neural network equal to the first Piola-Kirchhoff stress $\boldsymbol{P}_u\left(\boldsymbol{F}\left(\boldsymbol{X}, \boldsymbol{\phi}\right)\right)$ computed using the displacement predicted by the neural network: 
\begin{equation}\label{constitutiveP}
\begin{aligned}
    \boldsymbol{P}\left(\boldsymbol{X}, \boldsymbol{\phi}\right) &=  \boldsymbol{P}_u\left(\boldsymbol{F}\left(\boldsymbol{X}, \boldsymbol{\phi}\right)\right).\\
\end{aligned}
\end{equation}
The computation of the loss function and dependent variables customarily requires obtaining the derivatives of the solution, which are determined using the automatic differentiation available in deep learning frameworks. The unconstrained optimization (minimization) problem is expressed as:
\begin{equation}\label{Minimization}
\begin{aligned}
    \boldsymbol{\phi}^{*}&=\underset{\boldsymbol{\phi}}{\mathrm{arg\,min}} \; \mathcal{L}\left(\boldsymbol{\phi}\right){.}\\
\end{aligned}
\end{equation} 
The loss function we consider is written as:
\begin{equation}\label{loss_function}
\begin{split}
    \mathcal{L} &= \alpha_{\Pi} \Pi\left( \boldsymbol{u}\left(\left\{\boldsymbol{X}_{\Pi}^{i}\right\}_{i=1}^{N_{\Pi}}, \boldsymbol{\phi}\right)\right) + \alpha_{P} MSE_{P}\left(\left\{\boldsymbol{X}_{\Pi}^{i}\right\}_{i=1}^{N_{\Pi}}, \boldsymbol{\phi}\right) + \alpha^{u}_{t} MSE^{u}_{t}\left(\left\{\boldsymbol{X}_{t}^{i}\right\}_{i=1}^{N_{t}}, \boldsymbol{\phi}\right) \\ 
    & + \alpha^{P}_{t} MSE^{P}_{t}\left(\left\{\boldsymbol{X}_{t}^{i}\right\}_{i=1}^{N_{t}}, \boldsymbol{\phi}\right) + \alpha^{u}_{int} MSE^{u}_{int}\left(\left\{\boldsymbol{X}_{int}^{i}\right\}_{i=1}^{N_{int}}, \boldsymbol{\phi}\right) + \alpha^{P}_{int} MSE^{P}_{int}\left(\left\{\boldsymbol{X}_{int}^{i}\right\}_{i=1}^{N_{int}}, \boldsymbol{\phi}\right)\\
\end{split}
\end{equation}
where $\alpha_{\Pi}$, $\alpha_{P}$, $\alpha^{P}_{t}$, $\alpha^{u}_{t}$, $\alpha^{P}_{int}$ and $\alpha^{u}_{int}$ are the weight of the different loss terms computed using the CoV method discussed below. $\boldsymbol{X}_{\Pi}^{i}$ are domain training points, $\boldsymbol{X}_{int}^{i}$ denote the interior points of the domain, and $\boldsymbol{X}_{t}^{i}$ represent the traction boundary condition points $\overline{\boldsymbol{t}}$. The displacement boundary conditions are enforced using the distance functions highlighted earlier, so there is no need to account for them while defining our loss function. The first term in Equation (\ref{loss_function}) accounts for the weak form, and it is expressed as:
\begin{equation}\label{hyper_weak_loss}
\begin{split}
    \Pi\left(\left\{\boldsymbol{X}_{\Pi}^{i}\right\}_{i=1}^{N_{\Pi}}, \boldsymbol{\phi}\right) &= \int_{\Omega} \psi\left(\left\{\boldsymbol{X}_{\Pi}^{i}\right\}_{i=1}^{N_{\Pi}}, \boldsymbol{\phi}\right) d\Omega - \int_{\Omega} \boldsymbol{u}^{T}\left(\left\{\boldsymbol{X}_{\Pi}^{i}\right\}_{i=1}^{N_{\Pi}}, \boldsymbol{\phi}\right) \;\boldsymbol{f}_{b} d\Omega \\
    & - \int_{\Gamma} \boldsymbol{u}^{T}\left(\left\{\boldsymbol{X}_{\Pi}^{i}\right\}_{i=1}^{N_{\Pi}}, \boldsymbol{\phi}\right)\; \overline{\boldsymbol{t}} d\Gamma.\\
\end{split}
\end{equation}
The second term in Equation (\ref{loss_function}) accounts for the learning of the constitutive law, and it is expressed as:
\begin{equation}\label{constitutive_loss}
\begin{split}
    MSE_{P}\left(\left\{\boldsymbol{X}_{\Pi}^{i}\right\}_{i=1}^{N_{\Pi}}, \boldsymbol{\phi}\right) &= \frac{1}{N_{\Pi}} \sum_{i=1}^{N_{\Pi}}\Big\Vert \boldsymbol{P}\left(\left\{\boldsymbol{X}_{\Pi}^{i}\right\}_{i=1}^{N_{\Pi}}, \boldsymbol{\phi}\right) -  \boldsymbol{P}_u\left(\boldsymbol{F}\left(\left\{\boldsymbol{X}_{\Pi}^{i}\right\}_{i=1}^{N_{\Pi}}, \boldsymbol{\phi}\right)\right) \Big\Vert^2.\\
\end{split}
\end{equation}
The third and fourth terms in Equation (\ref{loss_function}) reckon the traction boundary conditions; the third term is based on $\boldsymbol{P}_u\left(\boldsymbol{F}\left(\left\{\boldsymbol{X}_{t}^{i}\right\}_{i=1}^{N_{t}}, \boldsymbol{\phi}\right)\right)$, while the fourth term is calculated using $\boldsymbol{P}\left(\left\{\boldsymbol{X}_{t}^{i}\right\}_{i=1}^{N_{t}}, \boldsymbol{\phi}\right)$:
\begin{equation}\label{traction_1}
\begin{aligned}
   MSE^{u}_{t}\left(\left\{\boldsymbol{X}_{t}^{i}\right\}_{i=1}^{N_{t}}, \boldsymbol{\phi}\right) &= \frac{1}{N_{t}} \sum_{i=1}^{N_{t}}\Big\Vert \boldsymbol{P}_u\left(\boldsymbol{F}\left(\left\{\boldsymbol{X}_{t}^{i}\right\}_{i=1}^{N_{t}}, \boldsymbol{\phi}\right)\right) - \overline{\boldsymbol{t}}\left(\left\{\boldsymbol{X}_{t}^{i}\right\}_{i=1}^{N_{t}}\right)\Big\Vert^2,\\
    MSE^{P}_{t}\left(\left\{\boldsymbol{X}_{t}^{i}\right\}_{i=1}^{N_{t}}, \boldsymbol{\phi}\right) &= \frac{1}{N_{t}} \sum_{i=1}^{N_{t}} \Big\Vert \boldsymbol{P}\left(\left\{\boldsymbol{X}_{t}^{i}\right\}_{i=1}^{N_{t}}, \boldsymbol{\phi}\right) - \overline{\boldsymbol{t}}\left(\left\{\boldsymbol{X}_{t}^{i}\right\}_{i=1}^{N_{t}}\right)\Big\Vert^2.\\
\end{aligned}
\end{equation}
The last two terms of the loss function, shown in Equation (\ref{loss_function}), account for the residuals of the PDEs in the interior of the physical domain, where $MSE^{u}_{int}\left(\left\{\boldsymbol{X}_{int}^{i}\right\}_{i=1}^{N_{int}}, \boldsymbol{\phi}\right)$ and  $MSE^{P}_{int}\left(\left\{\boldsymbol{X}_{int}^{i}\right\}_{i=1}^{N_{int}}, \boldsymbol{\phi}\right)$ are computed using $\boldsymbol{P}_u\left(\boldsymbol{F}\left(\left\{\boldsymbol{X}_{int}^{i}\right\}_{i=1}^{N_{int}}, \boldsymbol{\phi}\right)\right)$ and $\boldsymbol{P}\left(\left\{\boldsymbol{X}_{int}^{i}\right\}_{i=1}^{N_{int}}, \boldsymbol{\phi}\right)$, respectively:
\begin{equation}\label{divergence_1}
\begin{aligned}
   MSE^{u}_{int}\left(\left\{\boldsymbol{X}_{int}^{i}\right\}_{i=1}^{N_{int}}, \boldsymbol{\phi}\right) &= \frac{1}{N_{int}} \sum_{i=1}^{N_{int}}\Big\Vert \boldsymbol{\nabla}_{\boldsymbol{X}}\cdot \boldsymbol{P}_u\left(\boldsymbol{F}\left(\left\{\boldsymbol{X}_{int}^{i}\right\}_{i=1}^{N_{int}}, \boldsymbol{\phi}\right)\right) + \boldsymbol{f}_{B} \left(\left\{\boldsymbol{X}_{int}^{i}\right\}_{i=1}^{N_{int}}\right) \Big\Vert^2,\\
    MSE^{P}_{int}\left(\left\{\boldsymbol{X}_{int}^{i}\right\}_{i=1}^{N_{int}}, \boldsymbol{\phi}\right) &= \frac{1}{N_{int}} \sum_{i=1}^{N_{int}} \Big\Vert \boldsymbol{\nabla}_{\boldsymbol{X}}\cdot \boldsymbol{P}\left(\left\{\boldsymbol{X}_{int}^{i}\right\}_{i=1}^{N_{int}}, \boldsymbol{\phi}\right) + \boldsymbol{f}_{B} \left(\left\{\boldsymbol{X}_{int}^{i}\right\}_{i=1}^{N_{int}}\right) \Big\Vert^2.\\
\end{aligned}
\end{equation}

\subsection{CoV weighting}\label{CoV_Section}
The loss function is defined as a linear combination of loss terms, where every loss term quantifies the cost for an auxiliary objective or desired output. We use the CoV weighting scheme, discussed in the work of Groenendijk et al. \cite{groenendijk2021multi}, to find the (optimal) set of weights $\alpha_i$ for the combination of loss terms (see Equation (\ref{loss_function})). The core principle of the CoV weighting method revolves around that a loss term has been fulfilled when its variance has approached zero. This indicates that a loss term with a constant value should not be optimized further. Nevertheless, variance alone is not good enough since a loss term with a larger (mean) magnitude also has a higher absolute variance even if the loss is relatively less variant. Thus, the coefficient of variation $c_{i}$ is used:
\begin{equation}\label{CoVEq}
\begin{aligned}
    c_{i} &= \frac{\sigma_{i}}{\mu_{i}}\\
\end{aligned}
\end{equation}
where $\sigma_i$ and $\mu_i$ denote the standard deviation and mean, respectively. The coefficient of variation displays the variability of the observed loss with the observed mean, and it is independent of the scale on which the observations are computed. In other words, the coefficient of variation decouples the magnitude of a loss term from its weighting. For example, a loss term with a small magnitude is still relevant when it is variant, while a loss term with a large magnitude, but it barely varies throughout the training process, is assigned less weight. This is vital because the different loss terms act on very different scales. 

Loss ratio $l_{i}$ is used rather than the loss value itself:
\begin{equation}\label{loss_ratio}
\begin{aligned}
    l_{i}^{t} &= \frac{\mathcal{L}_{i}^{t}}{\mu_{\mathcal{L}_{i}^{t-1}}}\\
\end{aligned}
\end{equation}
where the subscript $i$ refers to the $i^{th}$ term of the loss function defined in Equation (\ref{loss_function}), and $t$ denotes the iteration number. $\mu_{\mathcal{L}_{i}^{t-1}}$ is the mean of the loss term $i$ up to iteration $t-1$. The weight $\alpha_{l_{i}^{t}}$ is calculated using the coefficient of variation of the loss ratio $c_{l_{i}^{t}}$ for the loss term $\mathcal{L}_{i}$ at iteration $t$:
\begin{equation}\label{alpha_weight}
\begin{aligned}
    \alpha_{l_{i}^{t}} &= \frac{1}{z_{t}} c_{l_{i}^{t}} = \frac{1}{z_{t}} \frac{\sigma_{l_{i}^{t}}}{\mu_{l_{i}^{t}}}\\
\end{aligned}
\end{equation}
where $z_t$ is a normalizing parameter independent of $i$:
\begin{equation}\label{z_t}
\begin{aligned}
    z_t &= \sum_{i=1} c_{l_{i}^{t}}.\\
\end{aligned}
\end{equation}
This enforces that $\sum_{i=1} \alpha_{l_{i}^{t}} = 1$. Welford’s algorithm \cite{welford1962note} is utilized to determine the coefficient of variation and loss ratio: 
\begin{equation}\label{Welford}
\begin{aligned}
    \mu_{\mathcal{L}_{i}^{t}} &= \left(1-\frac{1}{t}\right) \mu_{\mathcal{L}_{i}^{t-1}} + \frac{1}{t} \mathcal{L}_{i}^{t},\\
    \mu_{l_{i}^{t}} &= \left(1-\frac{1}{t}\right) \mu_{l_{i}^{t-1}} + \frac{1}{t} l_{i}^{t},\\
    M_{l_{i}^{t}} &= \left(1-\frac{1}{t}\right) M_{l_{i}^{t-1}} + \frac{1}{t} \left(l_{i}^{t} - \mu_{l_{i}^{t-1}}\right) \left(l_{i}^{t} - \mu_{l_{i}^{t}}\right), \quad \text{and}\\
    \sigma_{l_{i}^{t}} &= \sqrt{M_{l_{i}^{t}}}.\\
\end{aligned}
\end{equation}
By following the CoV weighing method, one infers that there are two forces that determine the weight of a loss term. 1) The loss weight increases when the loss ratio $l_{i}^{t}$ decreases, i.e., when the loss $\mathcal{L}_{i}^{t}$ is below the mean loss $\mu_{\mathcal{L}_{i}^{t}}$. This stimulates losses that are learning fast, and it dulls the influence of outliers on the magnitude of the loss. 2) The loss weight grows when the standard deviation increases. This guarantees that more learning transpires when the loss ratio is more variant.
\section{Numerical examples}\label{results}

This section covers different examples of how the proposed framework is used to solve 3D hyperelastic problems. The proposed framework is implemented in PyTorch \cite{NEURIPS2019_9015}, while the network parameters are optimized using the limited Broyden–Fletcher–Goldfarb–Shanno (LBFGS) algorithm with Strong Wolfe line search \cite{liu1989limited, lewis2013nonsmooth}. The numerical integrations, required for calculating $\Pi\left(\left\{\boldsymbol{X}_{\Pi}^{i}\right\}_{i=1}^{N_{\Pi}}, \boldsymbol{\phi}\right)$ defined in Equation (\ref{hyper_weak_loss}), are computed using the Simpson's rule. The examples discussed below include using the two hyperelastic models discussed earlier: Neo-Hookean and Lopez-Pamies hyperelastic models. Also, we consider scenarios of displacement- and traction-controlled application of load.  

High-throughput computations are performed on iForge, which is an HPC cluster hosted at the National Center for Supercomputing Applications (NCSA). A representative compute node on iForge has the following hardware specifications: two NVIDIA Tesla V100 GPUs, two $20$-core Intel Skylake CPUs, and $192$ GB main memory. The proposed PINN framework is a standalone approach. In other words, it does not need other numerical schemes or analytical solutions for data generation. However, we obtain reference solutions using finite element analyses for validation purposes only. We use the normalized ``$L_2$-error" as a metric reflecting on the accuracy of the PINN solution:
\begin{equation}\label{L2error}
\begin{aligned}
   L_2\text{-error}&=\frac{||\boldsymbol{u}_{REF}-\hat{\boldsymbol{u}}||^{L_{2}}}{||\boldsymbol{u}_{REF}||^{L_{2}}}{.}\\
\end{aligned}
\end{equation}

\subsection{Neo-Hookean cantilever beam}\label{NHCBTCL}

\begin{figure}[!htb]
    \centering
    \includegraphics[width=0.7\textwidth]{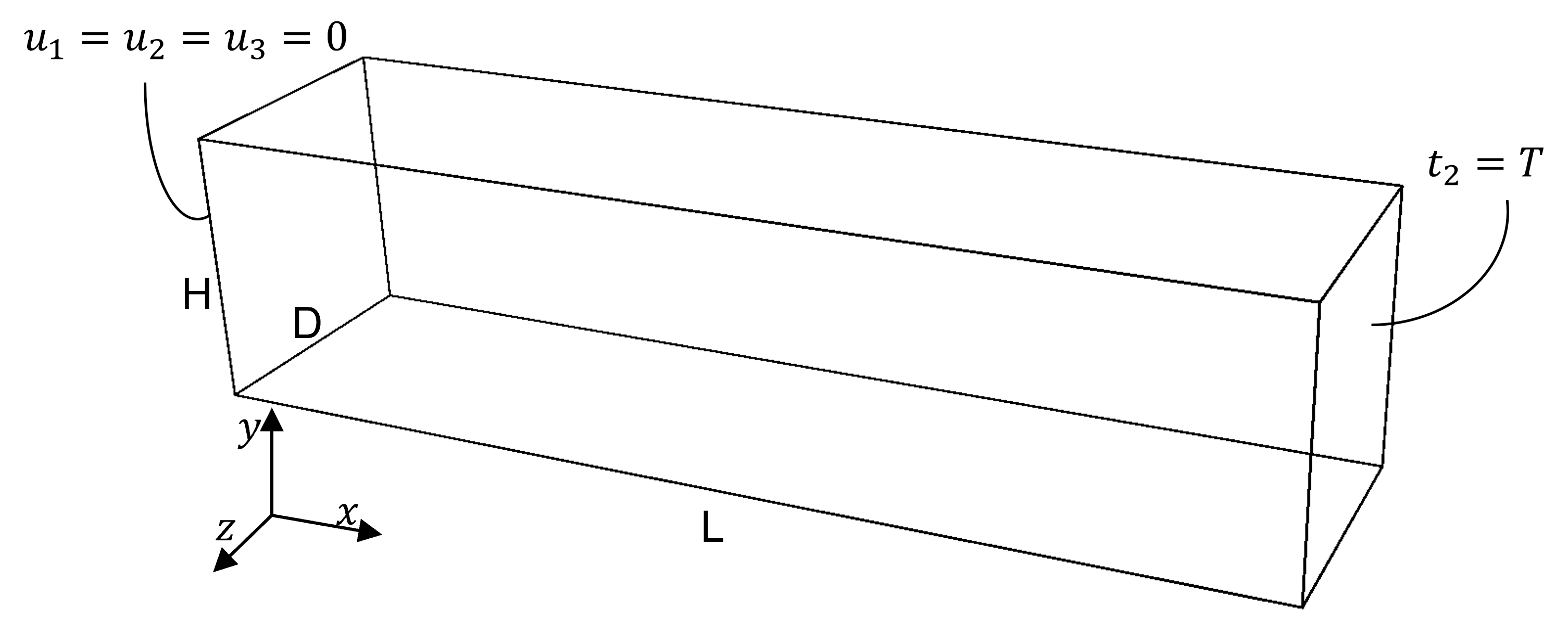}
    \caption{3D cantilever beam under nonzero traction.}
    \label{Example1}
\end{figure}

Let us consider a 3D neo-Hookean cantilever beam fixed at one end and under nonzero traction distributed at the other end, as illustrated in Figure \ref{Example1}. Let $L = 4\;\text{(m)}$ and $D=H=1\;\text{(m)}$. The traction is applied in the $y-$direction, with $T=-5$ (Pa). The material constansts are: $\lambda = 577$ (Pa) and $\mu = 385$ (Pa). Since this approach combines both the strong and weak forms, the degrees of freedom on the boundaries with both zero and nonzero tractions have to be explicitly satisfied when we define the loss terms related to the strong form. The $MSE^{u}_{t}\left(\left\{\boldsymbol{X}_{t}^{i}\right\}_{i=1}^{N_{t}}, \boldsymbol{\phi}\right)$ and $MSE^{P}_{t}\left(\left\{\boldsymbol{X}_{t}^{i}\right\}_{i=1}^{N_{t}}, \boldsymbol{\phi}\right)$ terms appearing in Equation (\ref{loss_function}) account for all traction boundary conditions needed for the strong form.

\begin{figure}[!htb]
    \centering
    \includegraphics[width=1.0\textwidth]{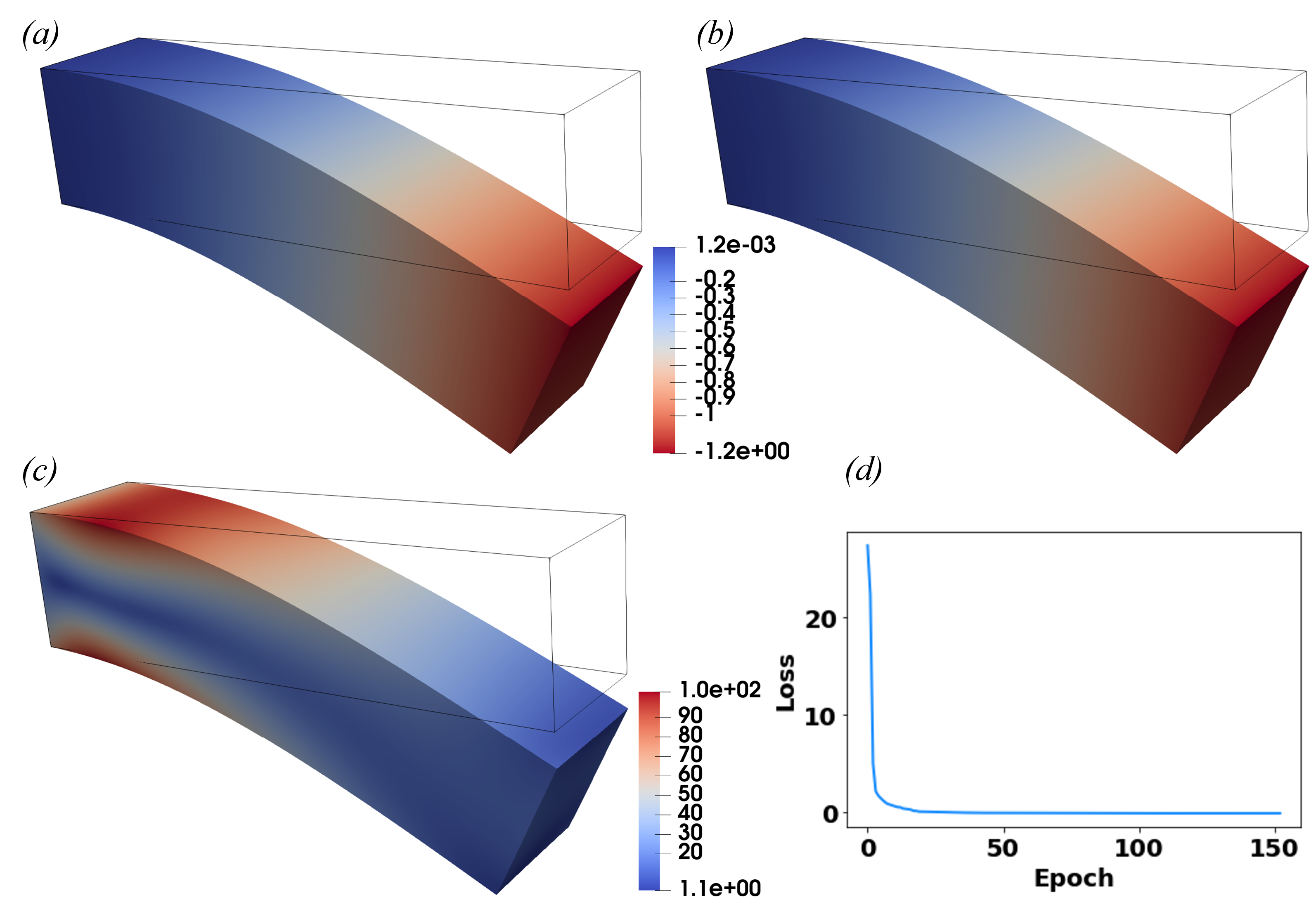}
    \caption{Neo-Hookean hyperelastic cantilever beam under nonzero traction: a) $u_{2}$ (m) obtained using the proposed framework, b) $u_{2}$ (m) obtained using the FEM, c) the von Mises stress (Pa), and d) total loss convergence history.}
    \label{NH_Neu_CB}
\end{figure}

Figure \ref{NH_Neu_CB} depicts the vertical displacement contours obtained from the proposed framework and compares them with those attained from the FEM. Additionally, the von Mises stress contours produced using the proposed PINN model are presented. When the solutions obtained from the FEM and PINN are compared, the $L_2\text{-error}$ is $0.034$. Figure \ref{NH_Neu_CB}d shows the convergence of the loss function. 

\subsection{Lopez-Pamies cantilever beam}\label{LPCBDCL}

\begin{figure}[!htb]
    \centering
    \includegraphics[width=0.7\textwidth]{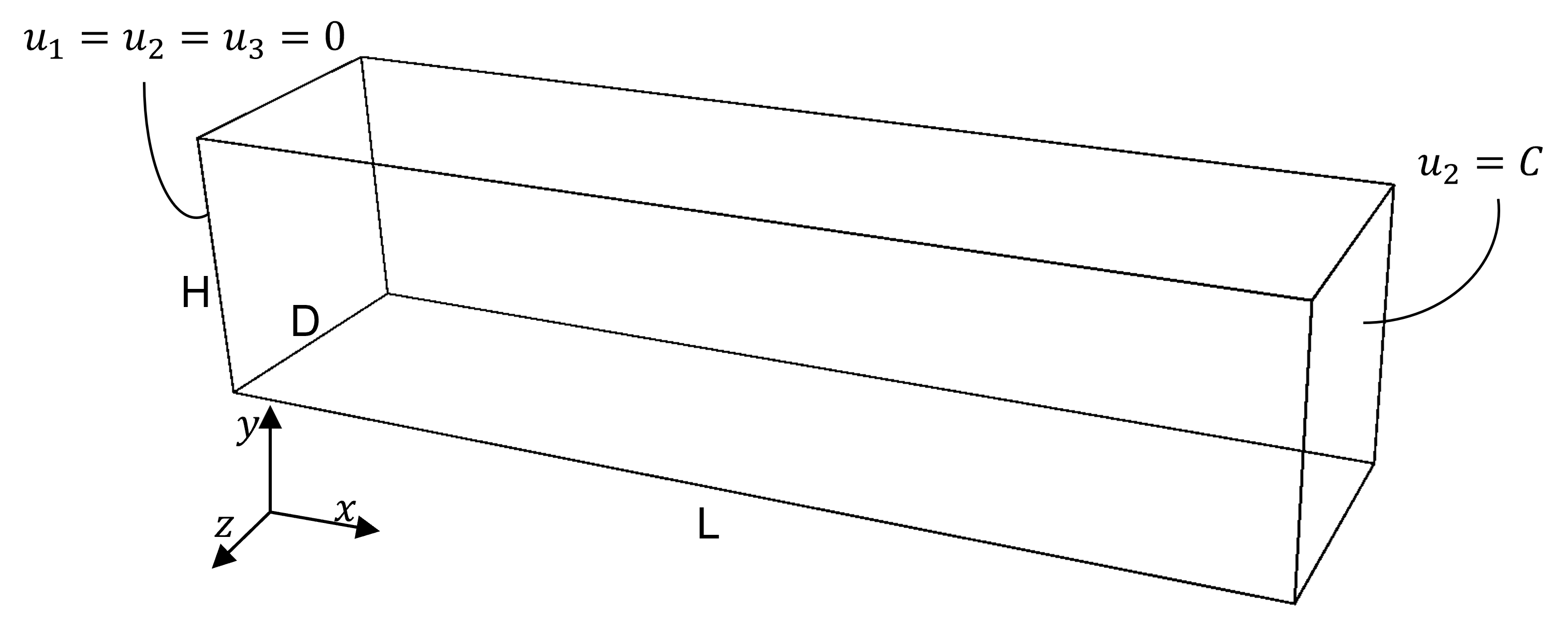}
    \caption{3D cantilever beam under nonzero displacement.}
    \label{Example2}
\end{figure}

\begin{figure}[!htb]
    \centering
    \includegraphics[width=1.0\textwidth]{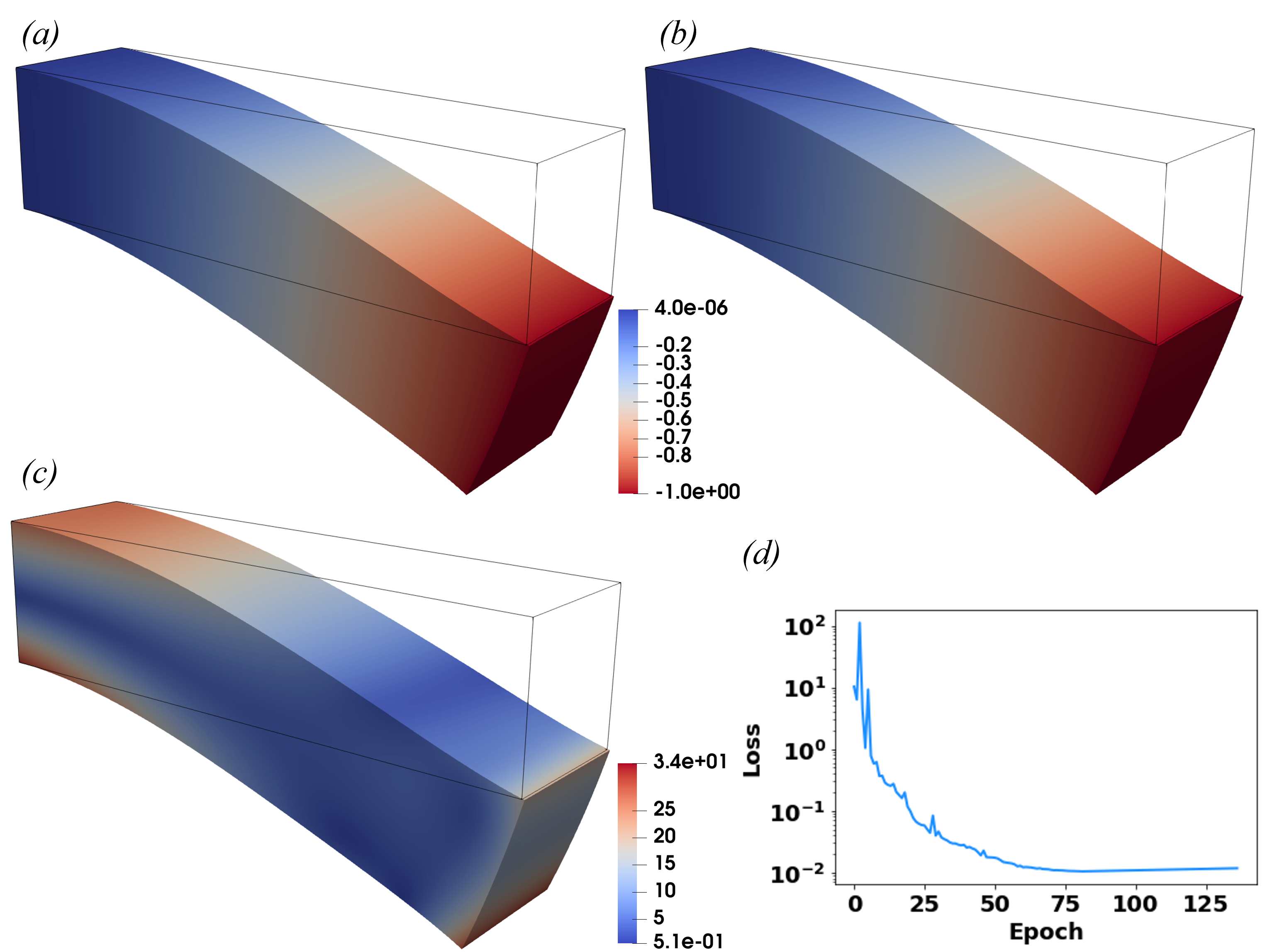}
    \caption{Lopez-Pamies hyperelastic cantilever beam under prescribed displacement: a) $u_{2}$ (m) obtained using the proposed framework, b) $u_{2}$ (m) obtained using the FEM, c) the von Mises stress (Pa), and d) total loss convergence history.}
    \label{OLP_Dir_CB}
\end{figure}

The second example we consider is a 3D Lopez-Pamies hyperelastic cantilever beam fixed at one end and under nonzero displacement at the other end, as depicted in Figure \ref{Example2}. Let $L = 4\;\text{(m)}$ and $D=H=1\;\text{(m)}$. The displacement is applied in the $y-$direction, with $C=-1$ (m). The material constants are defined as follows: $\alpha_1 =1$, $\mu_1 = 100$ (Pa), $\alpha_2 = -2$, $\mu_2=50$ (Pa), and $\lambda = 100$ (Pa). Figure \ref{OLP_Dir_CB} presents the vertical displacement contours obtained from the proposed framework and the FEM, as well as the von Mises stress contours. The $L_2\text{-error}$ is $0.0087$. Figure \ref{OLP_Dir_CB}d illustrates the convergence of the loss function. 

\subsection{Neo-Hookean cube under simple shear}\label{NHCSSDCL}

\begin{figure}[!htb]
    \centering
    \includegraphics[width=1.0\textwidth]{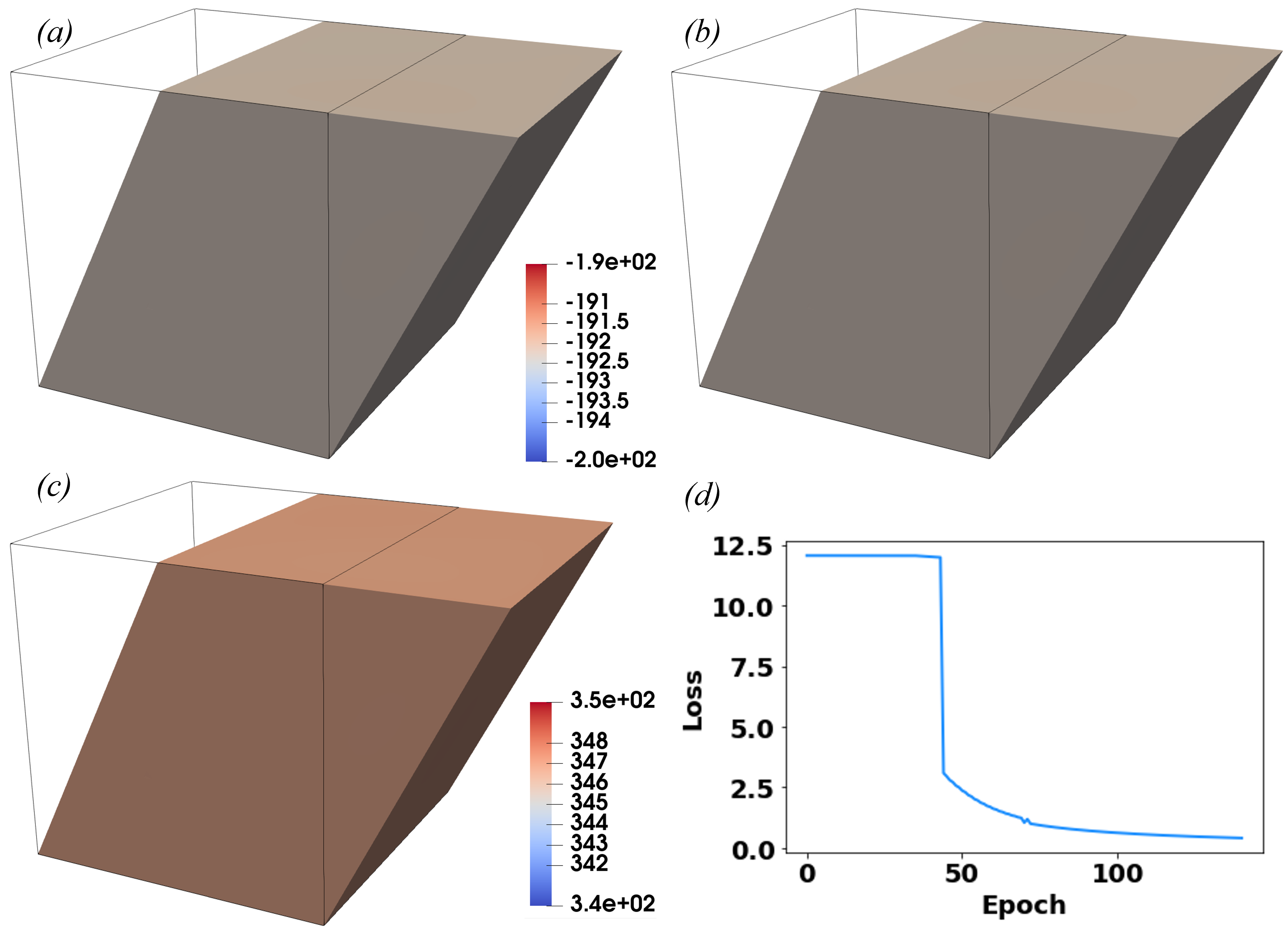}
    \caption{Neo-Hookean hyperelastic unit cube under simple shear: a) $S_{12}$ (Pa) obtained using the proposed framework, b) $S_{12}$ (Pa) obtained using the FEM, c) the von Mises stress (Pa), and d) total loss convergence history.}
    \label{NH_Dir_SS}
\end{figure}

Here, we consider a a neo-Hookean unit cube under simple shear. The solutions obtained from the proposed PINN model and FEA are found using a displacement-controlled scheme. The material constants used in this example are the same as those used in Section \ref{NHCBTCL}. Figure \ref{NH_Dir_SS} shows the Cauchy stress component $S_{12}$ obtained utilizing the PINN model and FEM, the von Mises stress, and the loss function's convergence history. The $L_2\text{-error}$ is $0.0065$.

\subsection{Neo-Hookean cube under localized traction}\label{NHCLTTCL}

\begin{figure}[!htb]
    \centering
    \includegraphics[width=0.5\textwidth]{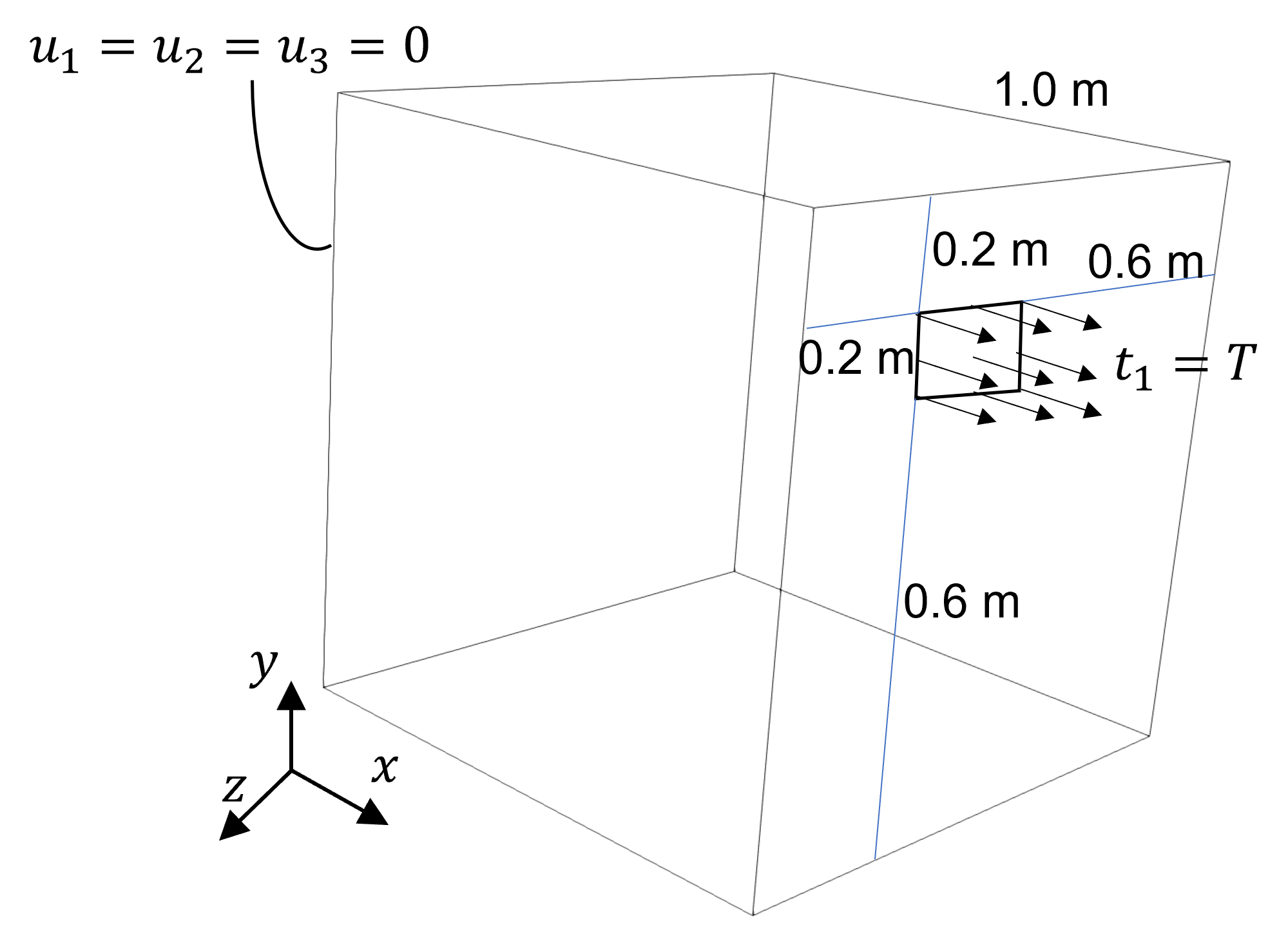}
    \caption{Unit cube under localized tensile traction.}
    \label{Example4}
\end{figure}

The last example we discuss is a neo-Hookean unit cube under localized tensile traction, as shown in Figure \ref{Example4}. The material parameters assigned in this example are the same as those used in Section \ref{NHCBTCL}. Zero displacements are imposed at the face opposite to the one with nonzero traction. The nonzero traction is applied on 4\% of the face area, while the remaining part has zero traction. The traction is applied in the $x-$direction, with $T=300$ (Pa).

\begin{figure}[!htb]
    \centering
    \includegraphics[width=1.0\textwidth]{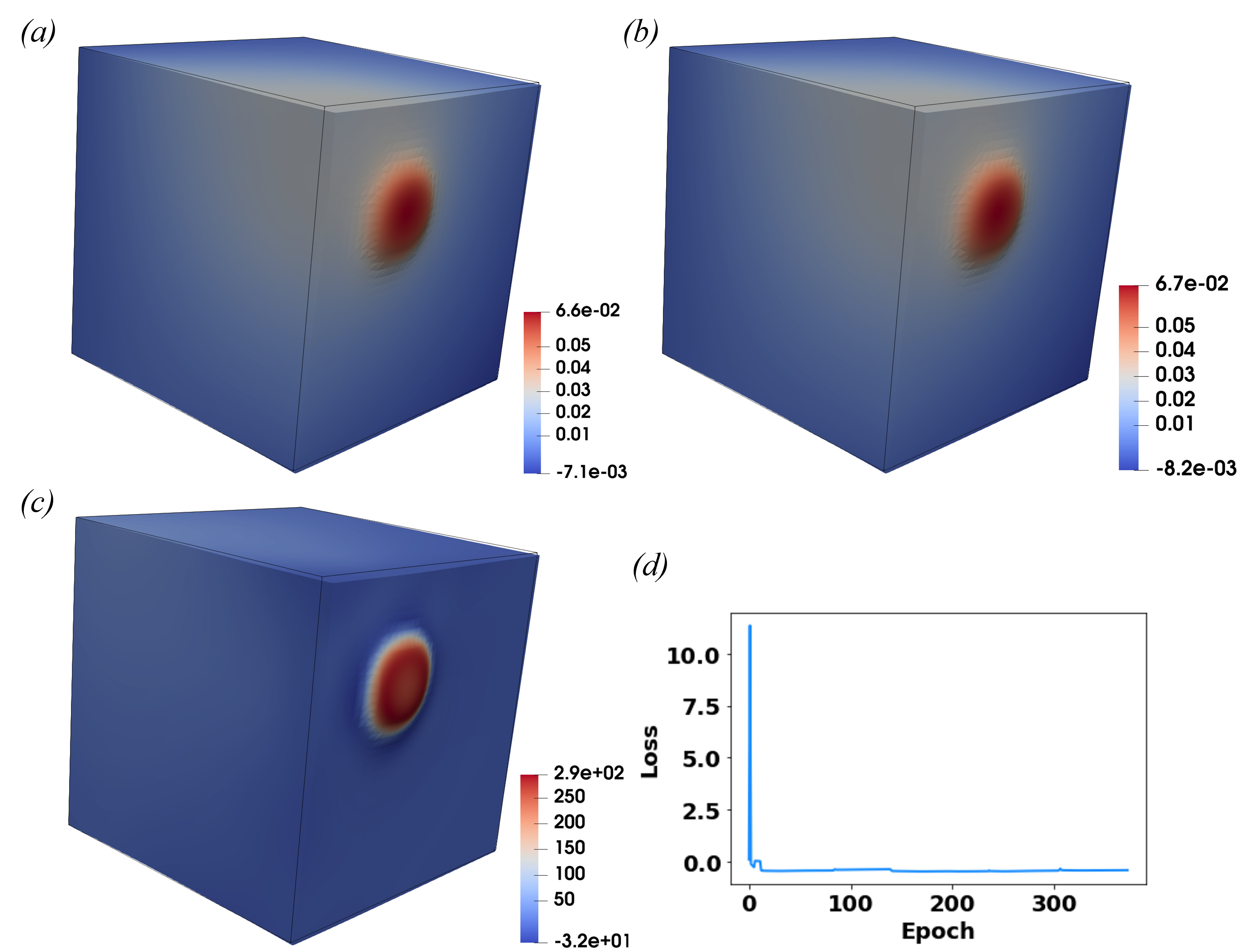}
    \caption{Neo-Hookean hyperelastic unit cube under localized traction: a) $u_{1}$ (m) obtained using the proposed framework, b) $u_{1}$ (m) obtained using the FEM, c) $S_{11}$ (Pa) obtained using the proposed framework, and d) total loss convergence history.}
    \label{NH_Neu_SCon}
\end{figure}

\begin{figure}[!htb]
    \centering
    \includegraphics[width=1.0\textwidth]{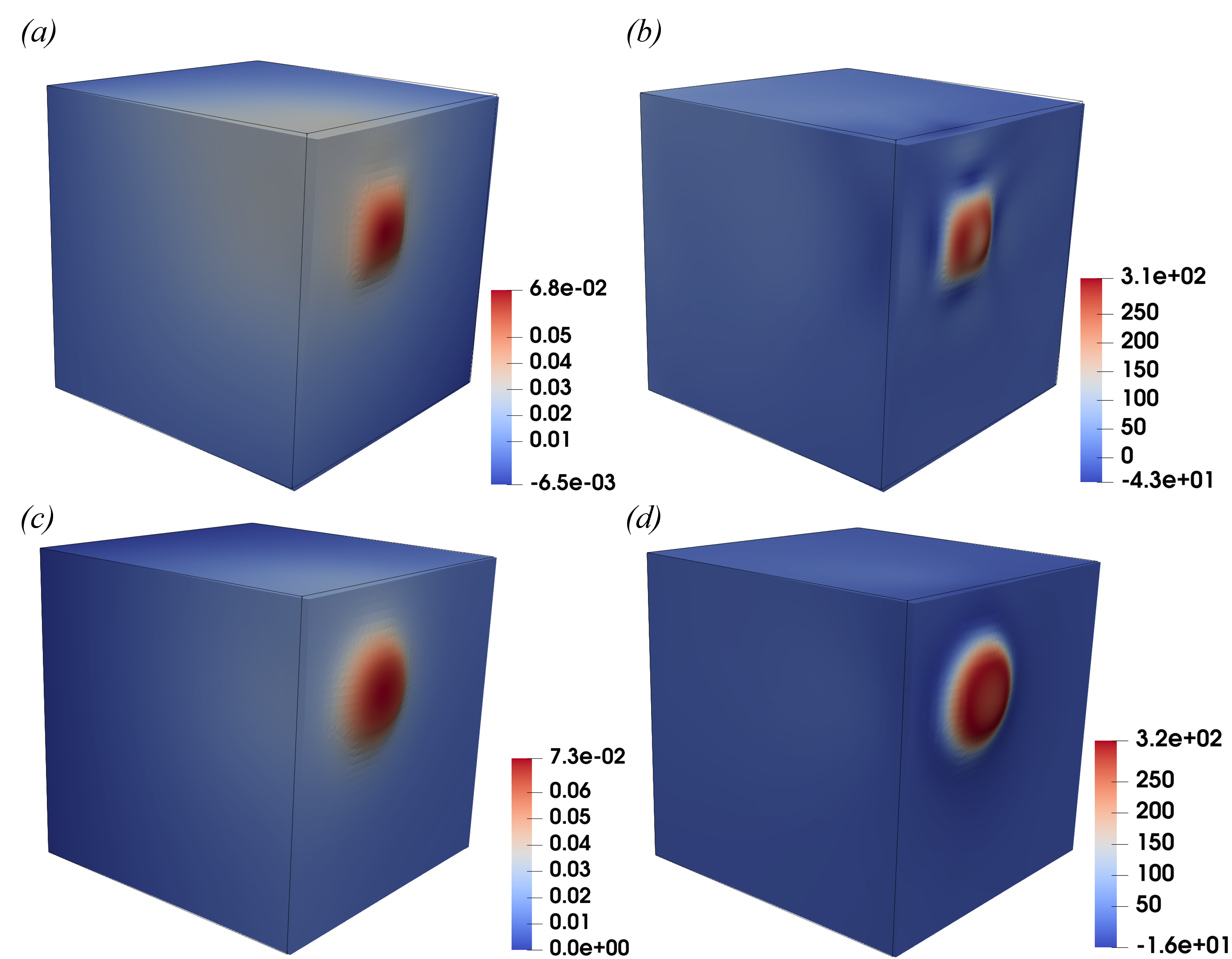}
    \caption{Neo-Hookean hyperelastic unit cube under localized traction: a) $u_{1}$ (m) obtained using the DEM, b) $S_{11}$ (Pa) obtained using the DEM, c) $u_{1}$ (m) obtained using the DCM, and d) $S_{11}$ (Pa) obtained using the DCM.}
    \label{DEM_DCM}
\end{figure}

Figure \ref{NH_Neu_SCon} depicts the displacement component $u_{1}$ obtained from the PINN model and FEM, the Cauchy stress component $S_{11}$, and the loss function's convergence history. The $L_2\text{-error}$ is $0.0091$. Also, we compare the solution obtained using the proposed PINN model with PINN models based on the seminal deep energy method (DEM) \cite{nguyen2020deep} and deep collocation method (DCM) \cite{abueidda2021meshless}. Figure \ref{DEM_DCM} shows the solutions obtained using the DEM and and DCM. The $L_2\text{-error}$ for the DEM and DCM are $0.034$ and $0.085$, respectively. The proposed framework outperforms both the DEM and DCM. 

\section{Discussion, conclusions, and future directions}\label{conclu}

This paper combines the potential energy, collocation method, and deep learning to solve partial differential equations governing the mechanical deformation of hyperelastic problems. Employing this approach does not necessitate any data generation, which is often the bottleneck in developing a data-driven model. The first pillar of the proposed framework is constructing the loss function. The loss function to be minimized consists of multi-loss terms, including the total potential energy, residuals of the strong form, and a term accounting for learning the constitutive relationship, as shown in Equation (\ref{loss_function}). The second pillar of the proposed PINN model is that the neural network demarcates the approximation space. Since the loss function has many loss terms, finding the suitable weights for the different loss terms is challenging. Therefore, we use the coefficient of variation weighting scheme to determine the weights. The proposed framework outperforms the seminal deep collocation and deep energy methods, especially when there are areas of high solution gradients, as shown in Section \ref{NHCLTTCL}. 

Moreover, the PINN model discussed in this paper is meshfree. Consequently, we neither define connectivity between the nodes nor involve mesh generation, which can be challenging in many cases \cite{bourantas2018strong} and may require partitioning methods for large meshes \cite{borrell2018parallel}. Also, meshfree methods avoid element distortion or volumetric locking. Typically, it is easier and quicker to write a code for a PINN model than to write a nonlinear finite element code from scratch. Also, when PINN models are used, transfer learning can be beneficial. Specifically, once the problem is solved for a particular load, the network parameters (weights and biases) can be stored to be used as the initial network parameters for a different problem if there are slight variations. Nonetheless, this is not the case for the classical numerical methods, where any variation in the problem requires solving the problem again from scratch.

The optimization problems used in determining the weights and biases of the PINN model are often nonconvex. Accordingly, one must be prudent of getting trapped in local minima. Nonconvexity yields several challenges that are to be investigated by the mechanics community. Also, in this work, the architecture of the network and hyperparameters are picked on a trial and error basis. Nevertheless, one needs an architecture and hyperparameters that provide good accuracy. Hamdia et al. \cite{hamdia2020efficient} proposed using the genetic algorithm to obtain the optimized architecture and hyperparameters, but not in the context of PINNs. We plan on showing the importance of selecting the hyperparameters and architecture of PINN models in a future paper, where we will scrutinize a systematic approach to optimizing the hyperparameters of a PINN model. Solving PDEs using PINNs is presently a growing trend within the research community, and this paper is by no means the final word on the topic.
\section*{Acknowledgment}
The authors would like to thank the National Center for Supercomputing Applications (NCSA) Industry Program and the Center for Artificial Intelligence Innovation.
\section*{Data availability}
The data that support the findings of this study are available from the corresponding author upon reasonable request. 

\bibliography{mybibfile}

\begin{thebibliography}{10}
\expandafter\ifx\csname url\endcsname\relax
  \def\url#1{\texttt{#1}}\fi
\expandafter\ifx\csname urlprefix\endcsname\relax\def\urlprefix{URL }\fi
\expandafter\ifx\csname href\endcsname\relax
  \def\href#1#2{#2} \def\path#1{#1}\fi

\bibitem{huerta2018meshfree}
A.~Huerta, T.~Belytschko, S.~Fern{\'a}ndez-M{\'e}ndez, T.~Rabczuk, X.~Zhuang,
  M.~Arroyo, Meshfree methods, Encyclopedia of Computational Mechanics Second
  Edition (2018) 1--38.

\bibitem{hughes2012finite}
T.~J. Hughes, The Finite Element Method: Linear Static and Dynamic Finite
  Element Analysis, Courier Corporation, 2012.

\bibitem{hughes2018isogeometric}
T.~J. Hughes, G.~Sangalli, M.~Tani, Isogeometric analysis: Mathematical and
  implementational aspects, with applications, in: Splines and PDEs: From
  Approximation Theory to Numerical Linear Algebra, Springer, 2018, pp.
  237--315.

\bibitem{kim2021deep}
Y.~Kim, Y.~Kim, C.~Yang, K.~Park, G.~X. Gu, S.~Ryu, Deep learning framework for
  material design space exploration using active transfer learning and data
  augmentation, npj Computational Materials 7~(1) (2021) 1--7.

\bibitem{shahane2022surrogate}
S.~Shahane, E.~Guleryuz, D.~W. Abueidda, A.~Lee, J.~Liu, X.~Yu, R.~Chiu,
  S.~Koric, N.~R. Aluru, P.~M. Ferreira, Surrogate neural network model for
  sensitivity analysis and uncertainty quantification of the mechanical
  behavior in the optical lens-barrel assembly, arXiv preprint
  arXiv:2201.09659.

\bibitem{rong2019predicting}
Q.~Rong, H.~Wei, X.~Huang, H.~Bao, Predicting the effective thermal
  conductivity of composites from cross sections images using deep learning
  methods, Composites Science and Technology 184 (2019) 107861.

\bibitem{mozaffar2019deep}
M.~Mozaffar, R.~Bostanabad, W.~Chen, K.~Ehmann, J.~Cao, M.~Bessa, Deep learning
  predicts path-dependent plasticity, Proceedings of the National Academy of
  Sciences 116~(52) (2019) 26414--26420.

\bibitem{koric2021deep}
S.~Koric, D.~W. Abueidda, Deep learning sequence methods in multiphysics
  modeling of steel solidification, Metals 11~(3) (2021) 494.

\bibitem{fatehi2021accelerated}
E.~Fatehi, H.~Y. Sarvestani, B.~Ashrafi, A.~Akbarzadeh, Accelerated design of
  architectured ceramics with tunable thermal resistance via a hybrid machine
  learning and finite element approach, Materials \& Design 210 (2021) 110056.

\bibitem{spear2018data}
A.~D. Spear, S.~R. Kalidindi, B.~Meredig, A.~Kontsos, J.-B. Le~Graverend,
  Data-driven materials investigations: \uppercase{t}he next frontier in
  understanding and predicting fatigue behavior, JOM 70~(7) (2018) 1143--1146.

\bibitem{he2022exploring}
J.~He, S.~Kushwaha, D.~Abueidda, I.~Jasiuk, Exploring the structure-property
  relations of thin-walled, 2d extruded lattices using neural networks, arXiv
  preprint arXiv:2205.06761.

\bibitem{gu2018bioinspired}
G.~X. Gu, C.-T. Chen, D.~J. Richmond, M.~J. Buehler, Bioinspired hierarchical
  composite design using machine learning: simulation, additive manufacturing,
  and experiment, Materials Horizons 5~(5) (2018) 939--945.

\bibitem{linka2021constitutive}
K.~Linka, M.~Hillg{\"a}rtner, K.~P. Abdolazizi, R.~C. Aydin, M.~Itskov, C.~J.
  Cyron, Constitutive artificial neural networks: a fast and general approach
  to predictive data-driven constitutive modeling by deep learning, Journal of
  Computational Physics 429 (2021) 110010.

\bibitem{raissi2019physics}
M.~Raissi, P.~Perdikaris, G.~E. Karniadakis, Physics-informed neural networks:
  A deep learning framework for solving forward and inverse problems involving
  nonlinear partial differential equations, Journal of Computational physics
  378 (2019) 686--707.

\bibitem{abueidda2021meshless}
D.~W. Abueidda, Q.~Lu, S.~Koric, Meshless physics-informed deep learning method
  for three-dimensional solid mechanics, International Journal for Numerical
  Methods in Engineering 122~(23) (2021) 7182--7201.

\bibitem{haghighat2021physics}
E.~Haghighat, M.~Raissi, A.~Moure, H.~Gomez, R.~Juanes, A physics-informed deep
  learning framework for inversion and surrogate modeling in solid mechanics,
  Computer Methods in Applied Mechanics and Engineering 379 (2021) 113741.

\bibitem{cai2021physics}
S.~Cai, Z.~Wang, S.~Wang, P.~Perdikaris, G.~E. Karniadakis, Physics-informed
  neural networks for heat transfer problems, Journal of Heat Transfer 143
  (2021) 6.

\bibitem{hornik1989multilayer}
K.~Hornik, M.~Stinchcombe, H.~White, Multilayer feedforward networks are
  universal approximators, Neural networks 2~(5) (1989) 359--366.

\bibitem{henkes2022physics}
A.~Henkes, H.~Wessels, R.~Mahnken, Physics informed neural networks for
  continuum micromechanics, Computer Methods in Applied Mechanics and
  Engineering 393 (2022) 114790.

\bibitem{niaki2021physics}
S.~A. Niaki, E.~Haghighat, T.~Campbell, A.~Poursartip, R.~Vaziri,
  Physics-informed neural network for modelling the thermochemical curing
  process of composite-tool systems during manufacture, Computer Methods in
  Applied Mechanics and Engineering 384 (2021) 113959.

\bibitem{rao2021physics}
C.~Rao, H.~Sun, Y.~Liu, Physics-informed deep learning for computational
  elastodynamics without labeled data, Journal of Engineering Mechanics 147~(8)
  (2021) 04021043.

\bibitem{samaniego2020energy}
E.~Samaniego, C.~Anitescu, S.~Goswami, V.~M. Nguyen-Thanh, H.~Guo, K.~Hamdia,
  X.~Zhuang, T.~Rabczuk, An energy approach to the solution of partial
  differential equations in computational mechanics via machine learning:
  Concepts, implementation and applications, Computer Methods in Applied
  Mechanics and Engineering 362 (2020) 112790.

\bibitem{nguyen2020deep}
V.~M. Nguyen-Thanh, X.~Zhuang, T.~Rabczuk, A deep energy method for finite
  deformation hyperelasticity, European Journal of Mechanics-A/Solids 80 (2020)
  103874.

\bibitem{nguyen2021parametric}
V.~M. Nguyen-Thanh, C.~Anitescu, N.~Alajlan, T.~Rabczuk, X.~Zhuang, Parametric
  deep energy approach for elasticity accounting for strain gradient effects,
  Computer Methods in Applied Mechanics and Engineering 386 (2021) 114096.

\bibitem{abueidda2022deep}
D.~W. Abueidda, S.~Koric, R.~A. Al-Rub, C.~M. Parrott, K.~A. James, N.~A. Sobh,
  A deep learning energy method for hyperelasticity and viscoelasticity,
  European Journal of Mechanics-A/Solids 95 (2022) 104639.

\bibitem{fuhg2022mixed}
J.~N. Fuhg, N.~Bouklas, The mixed deep energy method for resolving
  concentration features in finite strain hyperelasticity, Journal of
  Computational Physics 451 (2022) 110839.

\bibitem{krishnapriyan2021characterizing}
A.~Krishnapriyan, A.~Gholami, S.~Zhe, R.~Kirby, M.~W. Mahoney, Characterizing
  possible failure modes in physics-informed neural networks, in:
  A.~Beygelzimer, Y.~Dauphin, P.~Liang, J.~W. Vaughan (Eds.), Advances in
  Neural Information Processing Systems, 2021.

\bibitem{bengio2009curriculum}
Y.~Bengio, J.~Louradour, R.~Collobert, J.~Weston, Curriculum learning, in:
  Proceedings of the 26th annual international conference on machine learning,
  2009, pp. 41--48.

\bibitem{fuks2020limitations}
O.~Fuks, H.~A. Tchelepi, Limitations of physics informed machine learning for
  nonlinear two-phase transport in porous media, Journal of Machine Learning
  for Modeling and Computing 1~(1).

\bibitem{wang2021understanding}
S.~Wang, Y.~Teng, P.~Perdikaris, Understanding and mitigating gradient flow
  pathologies in physics-informed neural networks, SIAM Journal on Scientific
  Computing 43~(5) (2021) A3055--A3081.

\bibitem{wang2022and}
S.~Wang, X.~Yu, P.~Perdikaris, When and why pinns fail to train: A neural
  tangent kernel perspective, Journal of Computational Physics 449 (2022)
  110768.

\bibitem{tancik2020fourier}
M.~Tancik, P.~Srinivasan, B.~Mildenhall, S.~Fridovich-Keil, N.~Raghavan,
  U.~Singhal, R.~Ramamoorthi, J.~Barron, R.~Ng, Fourier features let networks
  learn high frequency functions in low dimensional domains, Advances in Neural
  Information Processing Systems 33 (2020) 7537--7547.

\bibitem{wang2021eigenvector}
S.~Wang, H.~Wang, P.~Perdikaris, On the eigenvector bias of \uppercase{F}ourier
  feature networks: From regression to solving multi-scale pdes with
  physics-informed neural networks, Computer Methods in Applied Mechanics and
  Engineering 384 (2021) 113938.

\bibitem{groenendijk2021multi}
R.~Groenendijk, S.~Karaoglu, T.~Gevers, T.~Mensink, Multi-loss weighting with
  coefficient of variations, in: Proceedings of the IEEE/CVF Winter Conference
  on Applications of Computer Vision, 2021, pp. 1469--1478.

\bibitem{chen2018gradnorm}
Z.~Chen, V.~Badrinarayanan, C.-Y. Lee, A.~Rabinovich, Gradnorm: Gradient
  normalization for adaptive loss balancing in deep multitask networks, in:
  International Conference on Machine Learning, PMLR, 2018, pp. 794--803.

\bibitem{lopez2010new}
O.~Lopez-Pamies, A new \uppercase{I}1-based hyperelastic model for rubber
  elastic materials, Comptes Rendus Mecanique 338~(1) (2010) 3--11.

\bibitem{koric2009explicit}
S.~Koric, L.~C. Hibbeler, B.~G. Thomas, Explicit coupled thermo-mechanical
  finite element model of steel solidification, International Journal for
  Numerical Methods in Engineering 78~(1) (2009) 1--31.

\bibitem{NEURIPS2019_9015}
A.~Paszke, S.~Gross, F.~Massa, A.~Lerer, J.~Bradbury, G.~Chanan, T.~Killeen,
  Z.~Lin, N.~Gimelshein, L.~Antiga, A.~Desmaison, A.~Kopf, E.~Yang, Z.~DeVito,
  M.~Raison, A.~Tejani, S.~Chilamkurthy, B.~Steiner, L.~Fang, J.~Bai,
  S.~Chintala, Pytorch: An imperative style, high-performance deep learning
  library, in: H.~Wallach, H.~Larochelle, A.~Beygelzimer, F.~d\textquotesingle
  Alch\'{e}-Buc, E.~Fox, R.~Garnett (Eds.), Advances in Neural Information
  Processing Systems 32, Curran Associates, Inc., 2019, pp. 8024--8035.

\bibitem{pattanayak2017pro}
S.~Pattanayak, Pattanayak, S.~John, Pro deep learning with tensorflow,
  Springer, 2017.

\bibitem{welford1962note}
B.~Welford, Note on a method for calculating corrected sums of squares and
  products, Technometrics 4~(3) (1962) 419--420.

\bibitem{liu1989limited}
D.~C. Liu, J.~Nocedal, On the limited memory \uppercase{BFGS} method for large
  scale optimization, Mathematical Programming 45~(1-3) (1989) 503--528.

\bibitem{lewis2013nonsmooth}
A.~S. Lewis, M.~L. Overton, Nonsmooth optimization via quasi-\uppercase{N}ewton
  methods, Mathematical Programming 141~(1-2) (2013) 135--163.

\bibitem{bourantas2018strong}
G.~Bourantas, G.~Joldes, A.~Wittek, K.~Miller, Strong-and weak-form meshless
  methods in computational biomechanics, in: Numerical Methods and Advanced
  Simulation in Biomechanics and Biological Processes, Elsevier, 2018, pp.
  325--339.

\bibitem{borrell2018parallel}
R.~Borrell, J.~C. Cajas, D.~Mira, A.~Taha, S.~Koric, M.~V{\'a}zquez,
  G.~Houzeaux, Parallel mesh partitioning based on space filling curves,
  Computers \& Fluids 173 (2018) 264--272.

\bibitem{hamdia2020efficient}
K.~M. Hamdia, X.~Zhuang, T.~Rabczuk, An efficient optimization approach for
  designing machine learning models based on genetic algorithm, Neural
  Computing and Applications (2020) 1--11.

\end{thebibliography}

\end{document}